\documentclass[a4paper]{jpconf}
\usepackage{latexsym}
\usepackage{graphicx}
\begin{document}
\title{Na-ion dynamics in Quasi-1D compound NaV$_2$O$_4$}

\author{Martin~M\aa nsson$^{1,2}$,
Izumi~Umegaki$^3$,
Hiroshi~Nozaki$^3$,
Yuki~Higuchi$^3$,
Ikuto~Kawasaki$^4$,
Isao~Watanabe$^4$,
Hiroya~Sakurai$^5$,
Jun~Sugiyama$^3$ }

\address{$^1$ Laboratory for Quantum Magnetism, $\acute{\rm E}$cole Polytechnique F$\acute{\rm e}$d$\acute{\rm e}$rale de Lausanne (EPFL),
CH-1015 Lausanne, Switzerland}
\address{$^2$ Laboratory for Neutron Scattering \& Imaging, Paul Scherrer Institute, CH-5232 Villigen PSI, Switzerland}
\address{$^3$ Toyota Central Research \& Development Laboratories, Inc., 41-1 Yokomichi, Nagakute, Aichi 480-1192, Japan}
\address{$^4$ Advanced Meson Science Laboratory, RIKEN, 2-1 Hirosawa, Wako, Saitama 351-0198, Japan}
\address{$^5$ National Institute for Materials Science, Namiki, Tsukuba, Ibaraki 305-0044, Japan}

\ead{martin.mansson@epfl.ch}

\begin{abstract}
We have used the pulsed muon source at ISIS to study high-temperature Na-ion dynamics in the quasi-one-dimensional (Q1D) metallic antiferromagnet NaV$_2$O$_4$. By performing systematic zero-field and longitudinal-field measurements as a function of temperature we clearly distinguish that the hopping rate increases exponentially above $T_{\rm diff}\approx250$~K. The data is well fitted to an Arrhenius type equation typical for a diffusion process, showing that the Na-ions starts to be mobile above $T_{\rm diff}$. Such results makes this compound very interesting for the tuning of Q1D magnetism using atomic-scale ion-texturing through the periodic potential from ordered Na-vacancies. Further, it also opens the door to possible use of NaV$_2$O$_4$ and related compounds in energy related applications.
\end{abstract}

\section{Introduction}
The interest for Quasi-one-dimensional (Q1D) magnets have been extensive in both the experimental, as well as theoretical research communities \cite{Matsuda,Mihaly,Zvyagin}. The reason is that these materials display a variety of fascinating phenomena, such as unconventional magnetic ground states \cite{Sugiyama_01} and quantum effects \cite{Simutis}. The physics behind these phenomena is governed by the strong spin-spin interaction along the 1D direction, together with a much weaker coupling along the other directions. Further, it is well known that an ideal 1D AF spin system does not show long-range ordering (LRO) above $T=0$~K due to strong quantum spin fluctuation, causing a series of unconventional phenomena. Also Q1D metals have been in the center of attention for scientists worldwide since decades \cite{Grioni_Review}, due to their tendency toward Fermi surface instabilities, $e.g.$ density waves, and for their ability to form non-Fermi-liquid ground states. At present day, big efforts are still being conducted to understand spurious Q1D phenomena and to theoretically predict the ground states. $e.g.$ the superconducting (non-BSC) state that occur when applying pressure to the $\beta$-Na$_{0.33}$V$_{2}$O$_{5}$ compound \cite{Yamauchi,Okazaki}. Another example is BaVS$_{3}$ which showed Luttinger liquid behavior in its photoemission spectra \cite{Nakamura}.

\begin{figure}[t]
  \begin{center}
    \includegraphics[keepaspectratio=true,width=130 mm]{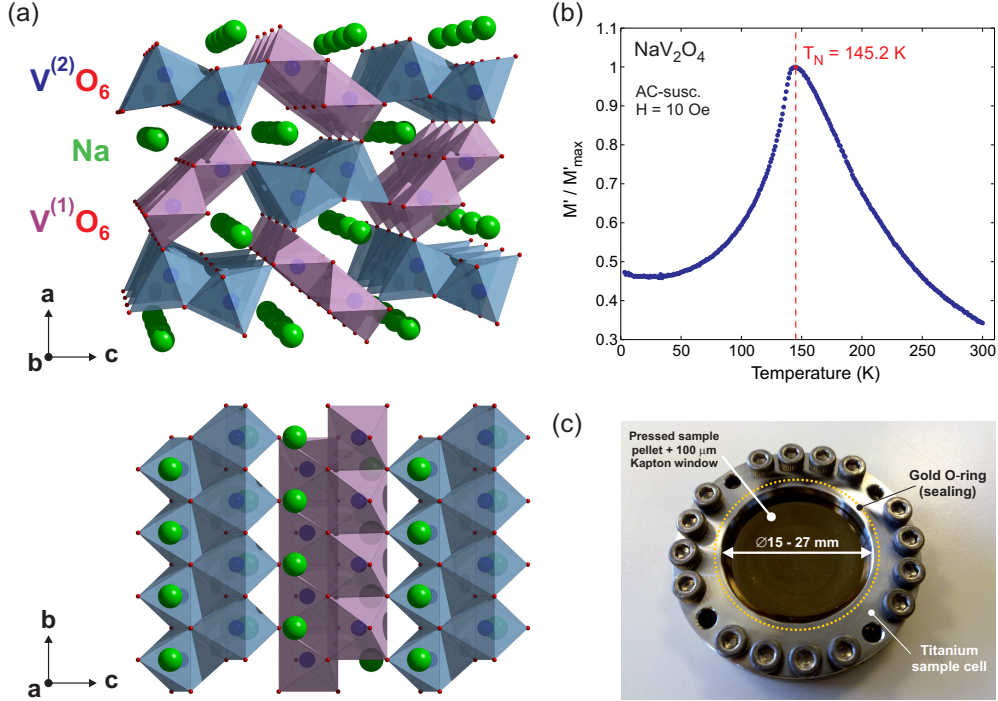}
  \end{center}
  \caption{(Color online)
  (a) Crystallographic structure of NaV$_2$O$_4$ for two different viewing directions that clearly shows the quasi-1D Na-ion diffusion channels along the \textbf{b}-axis.
  (b) Normalized AC-susceptibility showing how NaV$_2$O$_4$ enters into an antiferromagnetic long-range order below $T_{\rm N}\approx$~145.2~K.
  (c) Sample powder cell made of titanium showing the thin (50–100 $\mu$m) Kapton ’entrance window’ covering the pressed sample pellet. The yellow dashed line schematically show the underlying gold O-ring that is used for sealing the cell inside a helium glove-box to avoid sample degradation.
    }
  \label{fig:structure}
\end{figure}

Very recently, a new member of the Q1D vanadates, NaV$_2$O$_4$, where discovered and synthesized by our collaborators at the National Institute for Materials Science (NIMS) in Japan \cite{Sakurai} as well as by another group \cite{Yamaura}. This compound is very interesting since it is something as unusual as a Q1D metallic antiferromagnet. NaV$_2$O$_4$ belong to a CaFe$_2$O$_4$-type orthorhombic structure [see Fig. 1(a)] having a \emph{Pnma} space group. In this structure V$_2$O$_4$ double (\emph{zigzag}) chains are formed by a network of edge-sharing VO$_6$ octahedra align along the \textbf{b}-axis so as to make irregular hexagonal 1D channels. The sodium ions are located “in the centre” of such channels and are thought to be highly mobile in the direction along the \textbf{b}-axis. Our group has already studied NaV$_2$O$_4$ and related compounds extensively using both neutron scattering \cite{Nozaki} and muon spin rotation/relaxation ($\mu^{+}$SR) \cite{Sugiyama_02,Ofer}. Such studies, however, focused on clarifying their unconventional magnetic ground states that arise from the anisotropic Q1D spin interactions. However, in this brief report we instead wish to focus on the high-temperature ion-dynamics connected to the Q1D diffusion channels.

\section{Experimental Details}
A polycrystalline sample of NaV$_2$O$_4$ was prepared by a solid-state reaction technique under a pressure of 6 GPa using Na$_4$V$_2$O$_7$ and V$_2$O$_3$ powders as starting materials. A mixture of the two powders was packed in an Au capsule, then heated at 1300$^{\circ}$C for 1 h, and finally quenched to ambient $T$. A powder XRD analysis showed that the samples were single phase with an orthorhombic system of space group \emph{Pnma} at ambient $T$. DC-$\chi(T)$ data from the present sample reproduce the prior measurements showing the entrance into a AF state at $T_{\rm N}\approx$~145.2~K [see Fig.~1(b)]. The preparation and characterization of our samples were described in greater detail elsewhere \cite{Sakurai}.

For the $\mu^{+}$SR experiment, approximately 2 grams of NaV$_2$O$_4$ sample was pressed into a disc with a 24~mm diameter and 1.5~mm thickness. Inside a helium glove-box the disc was packed into a Au-sealed (gold O-ring) powder cell made of pure titanium using a thin ($100~\mu$m) Kapton film as 'entrance window' for the muons [see Fig~1(c)]. In addition, a silver mask with a hole matching the samples diameter was mounted onto the cell to ensure that the any minor background signal is non-relaxing in a wide temperature range. The cell was mounted onto the Cu end-plate of a liquid-He flow-type cryostat in the temperature range between 10 and 500~K. Subsequently, ZF-, weak transverse-field (wTF) and LF-$\mu^{+}$SR spectra were collected using the RIKEN-RAL / ARGUS spectrometer at the pulsed muon source ISIS/RAL in UK. The experimental techniques are described in more detail elsewhere \cite{muSR_book}.

\section{Results and Discussion}
In order to investigate the diffusive properties of Na-ions in the NaV$_2$O$_4$ compound we have used our previously presented $\mu^+$SR technique for studying solid state ion diffusion in battery related materials \cite{Sugiyama_03,Mansson}. In similarity to our previous extensive measurements of both Li-ion \cite{Sugiyama_03,Sugiyama_04,Sugiyama_05,Sugiyama_06,Sugiyama_07} and Na-ion \cite{Mansson} diffusion, a series of ZF, wTF~=~30~G as well as LF~=~5~G and 10~G $\mu^{+}$SR spectra were acquired in the temperature range between 140 and 500~K [see Fig.~2(a-b)]. At each temperature, the ZF and two LF spectra were fitted by an exponentially relaxing dynamic Kubo-Toyabe (KT) function plus a small background signal from the fraction of muons stopped mainly in the silver mask:
\begin{eqnarray}
 A_0\,P(t) &=&
  A_{\rm KT} G^{\rm DGKT}(\Delta, \nu, t)\exp(-\lambda t)+ A_{\rm BG}
\label{eq:DKT}
\end{eqnarray}
Furthermore, a global fitting procedure was employed over the entire temperature range using a common alpha ($\alpha$) and background asymmetry ($A_{\rm BG}$), but with temperature dependent field fluctuation rate ($\nu$) and relaxation rate ($\lambda$). The static width of the local field distribution ($\Delta$) was from individual fitting found to be more or less independent of temperature [$\Delta\approx0.2\cdot10^6$~s$^{-1}$] and was finally also treated as a common parameter for the global fit over the entire temperature range.

\begin{figure}[t]
  \begin{center}
    \includegraphics[keepaspectratio=true,width=155 mm]{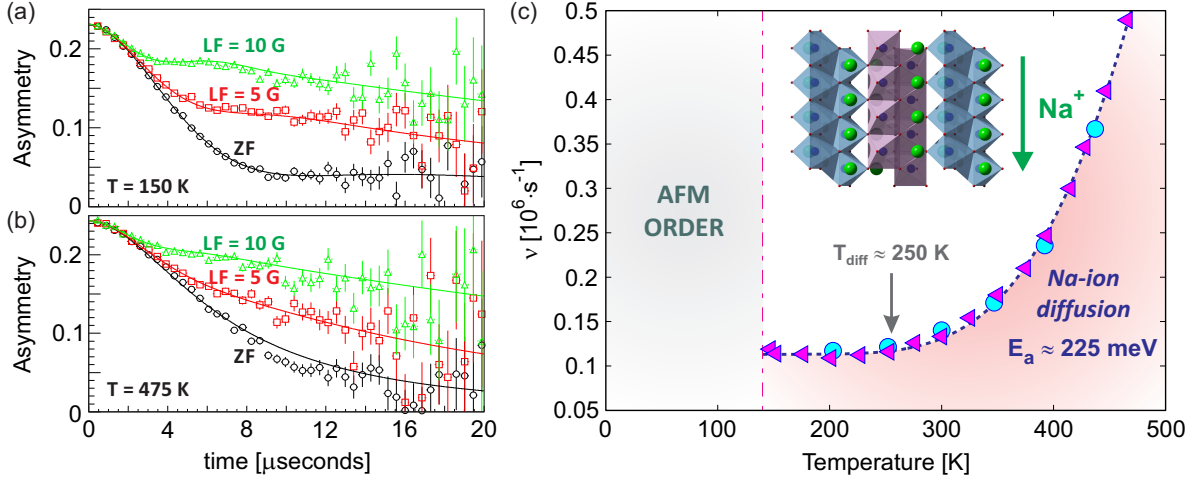}
  \end{center}
  \caption{(Color online)
  Zero-field (ZF) and Longitudinal-field (LF = 5 and 10 G) $\mu^+$SR time spectra collected at (a) $T=150$~K and (b) $T=475$~K.
  (c) Temperature dependence of the hopping-rate, $\nu(T)$ showing the exponential increase above $T_{\rm diff}\approx250$~K. Dashed line is fit to an Arrhenius type equation that yields the activation energy $E_{\rm a}\approx225$~meV for the Na-ion diffusion process. Filled circles and triangles, respectively, represent data from two different experiments and samples that were performed in order to verify the reproducibility of our results.
    }
  \label{fig:structure}
\end{figure}

Below ambient temperature, the spectra display a clear static behavior as shown for instance by the $T=150$~K spectra in Fig.~2(a) and with moderately increasing temperature the spectra remains more or less the same. However, just below ambient temperature a clear dynamic contribution sets in, as shown in the $T=475$~K spectra [Fig.~2(b)]. The temperature dependence of the Na-ion hopping rate [$\nu(T)$] obtained from fitting the data to Eq.~1 is shown in Fig.~2(c). As can be seen $\nu(T)$ displays a clear diffusive behavior above $T_{\rm diff}\approx250$~K. By fitting $\nu(T)$ to an Arrhenius type equation [dashed line in Fig.~2(c)], it is possible to extract the activation energy $E_{\rm a}\approx225$~meV.

Our results clearly show that the Na-ions starts to be mobile above $T_{\rm diff}$ in the NaV$_2$O$_4$ compound. In fact it is intriguing that Na-ions only become static just above the magnetic transition temperature $T_{\rm N}\approx$~145.2~K. Such situation is rather similar to the layered frustrated magnet Na$_{\rm x}$CoO$_2$ (NCO). NCO is a rather famous compound with an intriguing phase-diagram displaying superconducting \cite{Takada_03,Sugiyama_08}, magnetic \cite{Sugiyama_09,Foo_04,Bayrakci,Mendels}, as well as thermoelectric \cite{Terasaki,Lee} phases that strongly depend on the Na-content ($x$). It is well known that in NCO Na-vacancies can be ordered in many different 2D and 3D configurations \cite{Wang_09,Roger_07,Zhang_05} and also undergo order-disorder transitions \cite{Morris_09,Hinuma_08,Igarashi_08} at several different temperatures depending on the Na content ($x$). It was lately demonstrated that the electronic properties of the CoO$_{2}$ layers are not only governed by the number of conduction electrons on the triangular Co layers (i.e. $x$) but also the (periodic) Coulomb potential caused by (ordered) Na vacancies \cite{Roger_07}. In fact, Na ordering and its effect on the Fermi surface was found as the key factor that separates the Curie-Weiss metal (x~$>$~0.5) and Pauli metal (x~$<$~0.5) \cite{Bobroff_06}. The reconstruction of the Fermi surface with small Fermi-surface pockets caused by Na superstructures has later been demonstrated by the Shubnikov–de Haas (SdH) oscillation effect \cite{Balicas_08}.

Further, Schultze $et~al.$ proved by muon spin relaxation ($\mu^{+}$SR) and susceptibility measurements that not only the Na order but also the Na dynamics is of great importance \cite{Schulze_08}. They observed a new magnetic phase transition at $T~$=~8~K for $x=0.8$ samples, that could be connected to thermal history, i.e. Na rearrangement/diffusion, in a narrow temperature range around 200~K. This was the first clear proof that the sodium movement at high temperature could strongly influence the low-temperature magnetic properties. Later, other members of the same laboratory followed up this study with an $^{23}$Na NMR investigation of the same samples \cite{Weller_09}. Their results show a rapidly increasing mobility and diffusion of Na ions above 200~K, and finally at $T_{m}~$=~291~K, the Na layers melts and enters into a 2D liquid state. Above $T_{m}$, the NMR response is similar to what has been observed only in superionic conductors containing Na layers \cite{Villa_80}. Finally, we have within our collaboration very recently revealed new detailed information regarding subtle structural transitions that unlocks the diffusion pathways \cite{Medarde} as well as presented novel data on the Na-diffusion in NCO \cite{Mansson}. This has opened up intriguing possibilities for tuning fundamental physical properties in correlated electron systems by controlling the dynamic properties and through that the self-assembling structure on a sub-nanoscale \cite{McElroy_05,Zhang_06,Foo_04,Julien_08}.

From our current $\mu^+$SR data it seems very promising that a similar situation could be present also in NaV$_2$O$_4$. First of all, the low-temperature magnetic structure of NaV$_2$O$_4$ is Q1D, rather complex and there has been some questions concerning details of the incommensurate (IC) magnetic nature \cite{Nozaki,Ofer}. Further, the ion-dynamic region is in very close vicinity to the magnetic phase, in fact, much closer than for the above mentioned NCO system. This makes it very plausible that a tuning of the low-temperature Q1D incommensurate magnetic order as well as electronic properties can be achieved through a careful control of the Na-vacancy landscape. Currently there are no investigations available concerning Na-deficient Na$_{\rm x}$V$_2$O$_4$ samples. However, it would be of high interest for us to investigate how the Na-content ($x$) affects both the low-temperature magnetic properties as well as how this can be connected to the ion-dynamic phase at temperatures just above the magnetic transition.

Finally, the fact that Na-ions are mobile around room-temperature also opens the door to possible use of NaV$_2$O$_4$ and related compounds in next generation energy devices/applications. Such materials could not only function $e.g.$ as an ionic conductor but the direction of conduction would be strongly anisotropic as determined by the Q1D crystal structure. This is very similar to the LiFePO$_4$ battery compound that presently is one of the most promising and investigated cathode materials \cite{Sugiyama_05,Wang_11}. However, further investigations of Na-deficient samples [Na$_{\rm x}$V$_2$O$_4$] are needed using both electrochemical methods as well as miscroscopic studies of ion-diffusion and changes in the atomic structure with $x$.

\section{Acknowledgments}
We are grateful to the staff of ISIS and RIKEN-RAL for their assistance with our $\mu^+$SR experiments. This research was financially supported by the Swiss National Science Foundation (SNSF), MEXT KAKENHI Grant No. 23108003 and JSPS KAKENHI Grant No. 26286084.

\section*{References}

\end{document}